\documentclass[twocolumn,amsmath,amstex,amssymb,longbibliography,floatfix,superscriptaddress,a4paper,prl]{revtex4-1}

\usepackage{hyperref}
\hypersetup{
	colorlinks=true,
	linkcolor=blue,
	filecolor=magenta,      
	urlcolor=cyan,
	pdftitle={Tuning between continuous time crystals and many-body scars in long-range XYZ spin chains}
}
\usepackage{float}
\usepackage{graphicx}
\usepackage{framed}
\usepackage{color}
\usepackage{xcolor}
\usepackage{dsfont}
\usepackage{afterpage}	
\usepackage{braket}
\usepackage{mathtools}
\usepackage{enumerate}
\usepackage{comment}
\usepackage{subfigure}

\definecolor{darkblue}{rgb}{0, 0, 0.8}

\definecolor{imcolor}{rgb}{0.5,0.,0.5}

\begin{document}

\title{Tuning between continuous time crystals and many-body scars in long-range XYZ spin chains}

\begin{abstract}
{
Persistent oscillatory dynamics in non-equilibrium many-body systems is a tantalizing manifestation of ergodicity breakdown that continues to attract much attention. Recent works have focused on two classes of such systems: discrete time crystals and quantum many-body scars (QMBS). While both systems host oscillatory dynamics, its origin is expected to be fundamentally different:  discrete time crystal is a phase of matter which spontaneously breaks the $\mathbb{Z}_2$ symmetry of the external periodic drive, while QMBS span a subspace of non-thermalizing eigenstates forming an su(2) algebra representation. Here we ask a basic question: is there a physical system that allows to tune between these two dynamical phenomena? In contrast to much previous work, we investigate the possibility of a \emph{continuous} time crystal (CTC) in undriven, energy-conserving systems exhibiting prethermalization. We introduce a long-range XYZ spin model and show that it encompasses both a CTC phase as well as QMBS. We map out the dynamical phase diagram using numerical simulations based on exact diagonalization and time-dependent variational principle in the thermodynamic limit. We identify a regime where QMBS and CTC order co-exist, and we discuss experimental protocols that reveal their similarities as well as key differences. 
}
\end{abstract}

\author{Kieran Bull}
\affiliation{School of Physics and Astronomy, University of Leeds, Leeds LS2 9JT, United Kingdom}
\author{Andrew Hallam}
\affiliation{School of Physics and Astronomy, University of Leeds, Leeds LS2 9JT, United Kingdom}
\author{Zlatko Papi\'c}
\affiliation{School of Physics and Astronomy, University of Leeds, Leeds LS2 9JT, United Kingdom}
\author{Ivar Martin}
\affiliation{Material Science Division, Argonne National Laboratory, Argonne, IL 08540, USA}

\maketitle

{ \bf \em Introduction.---} The basic tenet of thermodynamics is that when a substance contains many constituents, its macroscopic behavior can be efficiently described by just a few variables such as pressure, volume, and temperature. Microscopic details typically only enter the mechanism of dissipation, which accounts for the energy transfer from the large to the microscopic scale (heating).  As a rule of thumb, the higher the temperature, the faster the relaxation of any non-generic state that possesses some ordering, such as magnetization, unless the latter is explicitly conserved by the system's Hamiltonian. 

It thus came as a surprise when Rydberg atom experiments~\cite{Bernien2017} revealed long-lived oscillations of an order parameter in a very high energy density initial state.  The oscillations were subsequently understood  to be due to quantum many-body scars (QMBSs): a dynamically-decoupled subspace within the many-body Hilbert space, spanned by non-thermalizing eigenstates, which are not protected by a symmetry~\cite{Serbyn2021,MoudgalyaReview}. The Rydberg atom system evades the generic expectations for rapid relaxation stated above as QMBS form ``towers" with (nearly) equidistant energy spacings. Superpositions of tower states undergo periodic evolution, thus avoiding the dephasing that afflicts generic states. These QMBS towers can be understood semiclassically~\cite{Turner2017,wenwei18TDVPscar,Choi2018,Bull2020,Turner2020}, based on an analogy with quantum scars of a single particle in a stadium billiard~\cite{Heller84}. Importantly, this behavior was shown to occur also in higher dimensions~\cite{Michailidis2D,Hsieh2020,Dolev2020} and in the presence of certain kinds of perturbations~\cite{TurnerPRB, Khemani2018, Lin2020} including disorder~\cite{MondragonShem2020}. 
More generally, QMBS subspaces are now understood to originate from a (restricted) spectrum generating algebra (RSGA)~\cite{BernevigEnt, MotrunichTowers, Dea2020}, which has been shown to arise in a number of non-integrable lattice models~\cite{BernevigEnt, ShiraishiMori, Iadecola2019_2, Buca2019, Bull2019, OnsagerScars, MoudgalyaFendley, Pakrouski2020, Ren2020, Surace2020, Kuno2020}.  
As the system remains non-integrable, this class of phenomena represents a \emph{weak} violation of the Eigenstate Thermalization Hypothesis (ETH)~\cite{DeutschETH, SrednickiETH}.

A seemingly distinct way of evading the ETH is the formation of a continuous time crystal (CTC)~\cite{Else2017}.  In a CTC phase, the system is in a prethermal state that corresponds to a near-ground state in the rotating frame, while being at a very high energy-density in the lab frame~\cite{Abanin2017}. Being at a low temperature in the rotating frame, the system has an option of developing an order parameter, thus spontaneously breaking symmetry which may be unique to the rotating frame. 
Eventually, the system is expected to fully thermalize; however, if both the pretermalization time scale 
and the thermalization time scale  (corresponding to full equilibration in the lab frame)  increase with the system size, the result would be a long lived -- quasistatic -- ordering in the rotating frame, manifesting as a ``rotating" order parameter in the lab frame.

In this paper we address the question: are CTC and QMBS distinct mechanisms of ETH breaking? The two \emph{a priori} appear different: QMBSs reveal themselves for very special initial states, while CTC, being a phase of matter, is supposed to be characterized by an order parameter, with the same order parameter configuration (defined down to physically small but microscopically large volume) possibly originating from  very different microscopic states. Nevertheless, one might wonder if underlying the CTC there are scar-like towers of states that violate the ETH.  Below we introduce a long-range XYZ spin model, experimentally motivated by  systems of trapped ions and polar molecules, which realizes \emph{both} QMBS as well as CTC route for evading the ETH.  For sufficiently long-range interactions, our simulations using infinite matrix product state methods reveal signatures of spontaneous symmetry breaking in the thermodynamic limit and the formation of CTC. For weakly anisotropic couplings and irrespective of interaction range, we demonstrate the existence of QMBS. The phase diagram shows that even though there are regimes where CTC and QMBS co-exist, these are two distinct ph enomena. We discuss experimental protocols that can distinguish between them.

\begin{figure*}
    \centering
     \includegraphics[width=1\textwidth]{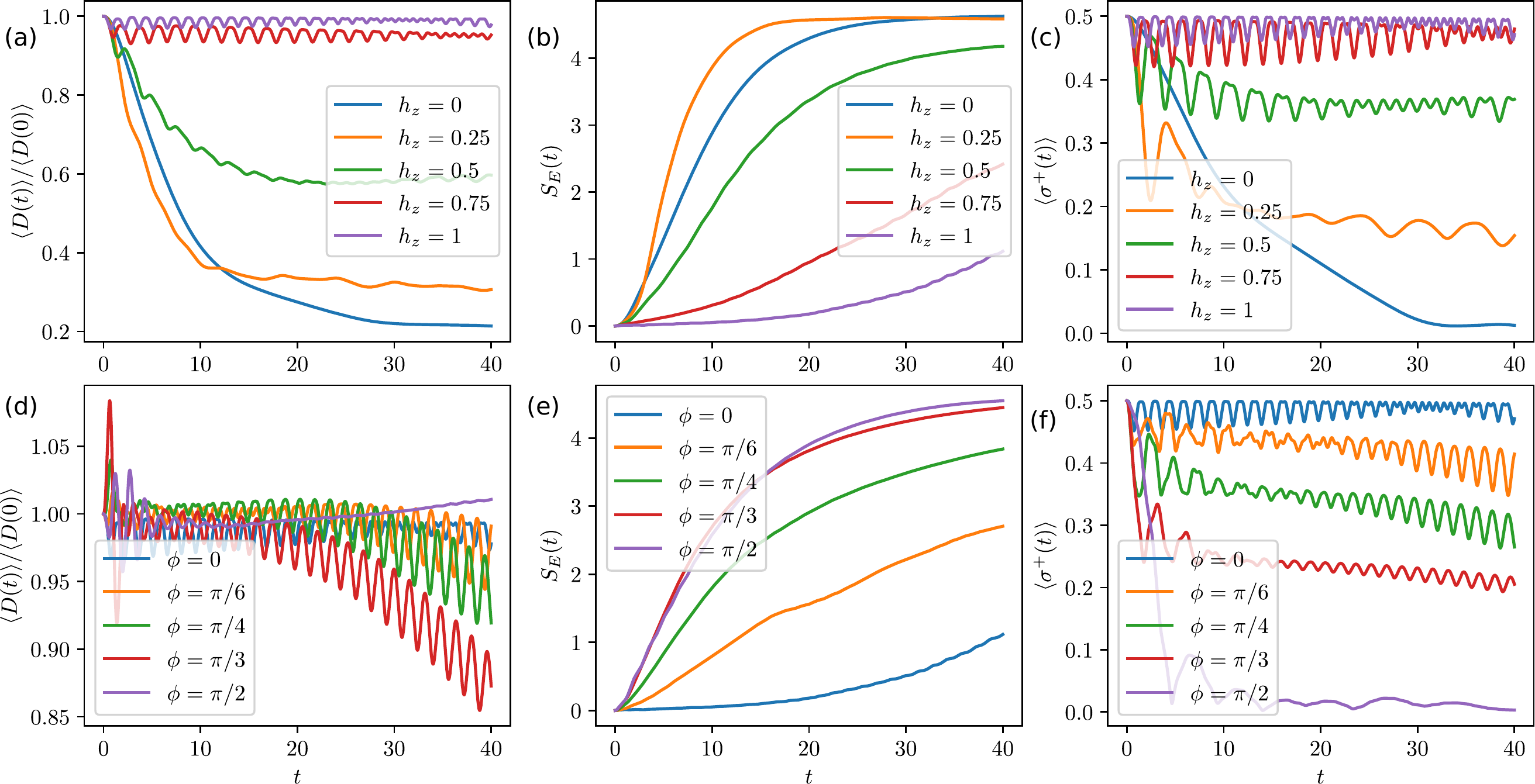}
    \caption{
    Signatures of a continuous time crystal. (a) Expectation value of the prethermal Hamiltonian, 
    Eq.~(\ref{eq:D}), in the time evolved state. The quantity is normalized by its value at time $t{=}0$. (b) Order parameter $\langle \sigma^+(t) \rangle$ defined in the text. (c) Entanglement entropy $S_E(t)$. 
    All plots are for the infinite long-range XYZ model in Eq.~(\ref{eq:Hamiltonian}) with  $\alpha=1.13$, $J_{x}{=}-0.4$, $J_{y}{=}-2.0$, $J_{z}{=}-1$. The value of the field $h_z$ is indicated in panels (a), (b), (c), while $h_z{=}1$ in panels (d), (e) and (f). The initial state is given by Eq.~(\ref{eq:initial}) with $\phi{=}0$ in (a), (b), (c), and $\phi=\{ 0,\pi/6, \pi/4, \pi/3, \pi/2 \}$ in panels (d), (e) and (f).
     }
    \label{fig:TC_data}
\end{figure*}

{ \bf \em The model.---} 
We focus on disorder-free systems, in which a discrete version of a time crystal (DTC) with spontaneously broken $\mathbb{Z}_2$ Ising symmetry has been demonstrated in numerics~\cite{Machado2020} and experiment~\cite{Kyprianidis2021}.  
By contrast, we consider an undriven XYZ spin model, with anisotropic long-range couplings and in the magnetic field along the $z$-axis, given by the Hamiltonian:
\begin{eqnarray}
H &=& \frac{1}{\mathcal{N}} \sum_{i>j}\sum_{\nu=x,y,z} \frac{J_\nu }{|i-j|^\alpha} \sigma_i^\nu \sigma_j^\nu  + h_z \sum_i \sigma_i^z,
\label{eq:Hamiltonian}
\end{eqnarray}
where $\sigma_i^\nu$ are the standard Pauli matrices on site $i$, and $\alpha$ controls the power-law decay of the interactions. 
We assume a 1D chain with open boundary conditions and  divide the interaction couplings $J^\nu$ with the Kac norm, $\mathcal{N}$,~\cite{SOM} which ensures the energy density is intensive.

Before we study the model in Eq.~(\ref{eq:Hamiltonian}) for $\alpha{>}1$, we note that the $\alpha{=}0$ limit is a fully connected  Lipkin-Meshkov-Glick (LMG)  model~\cite{LMG1,LMG2}, which
can be  described by only a few collective variables if initial states satisfy permutation symmetry~\cite{Sciolla2010,Sciolla2013,LerosePappalardi}. 
The  paramagnetic state ($J_\nu {\ll} |h_z|$) can be identified with a CTC (Ref.~\cite{Else2019Review} used a term ``mean-field time crystal" to distinguish this special type of CTC). 
In the Ising limit, the thermalization time  
is estimated as $\tau_\mathrm{th} {\sim} N^{\beta/d}$, where $\beta {=} \min(d{-}\alpha, (1{+}d)/2)$~\cite{Mori_2019}. Thus for $d{=}1$ and $\alpha {<} 1/2$, thermalization time (${\sim} N^{1{-}\alpha}$) is much longer than prethermalization or order parameter melting time, ${\sim}N^{1/2}$~\cite{Anderson1952,Tasaki2019}, both diverging with system size.

{ \bf \em Continuous time crystal.---} 
The possibility of a CTC in generic models away from the LMG limit was raised in Ref.~\cite{Else2017}. When $h_z\gg J_\nu$ and for short-range interactions, there is a U(1) charge, well-conserved over exponentially long times~\cite{Kuwahara2016,Abanin2017}.
The conditions for breaking  continuous U(1) symmetry  are  stringent: for short-range interactions, true long range order is only possible in $d{\ge}3$ (a classical Kosterlitz-Thouless transition can occur in $d{=}2$). In $d{=}1$, a phase transition requires long-range interactions.  
Our investigation below will focus on $\alpha>d{=}1$, for which bounds derived in Ref.~\cite{Machado2020} imply the existence of energy prethermalization.

For large $h_z$, sufficiently long-range interactions and low temperatures, Fig.~\ref{fig:TC_data} shows that a prethermal CTC phase emerges in the model given by Eq.~(\ref{eq:Hamiltonian}). Provided $h_z$ is sufficiently large, the dynamics in the model is described -- up to a timescale exponential in  $h_z/|J_x-J_y|$  -- by an effective Hamiltonian $H_\mathrm{eff}= D + h_z \sum_i \sigma_i^z$ where $D$ is given by~\cite{Else2017}
\begin{equation}
    D=\sum_{j>i}  \frac{1}{2}(J_x + J_y) \left(\frac{\sigma_i^x \sigma_j^x}{\vert i-j \vert^{\alpha}} + \frac{\sigma_i^y \sigma_j^y}{\vert i-j \vert^{\alpha}}\right) + J_z \frac{\sigma_i^z \sigma_j^z}{\vert i-j \vert^{\alpha}}.  \label{eq:D}
\end{equation}
This effective Hamiltonian has an emergent U(1) symmetry which is spontaneously broken at low (effective) temperatures for long-range interactions ($\alpha {\lesssim} 2.5$)~\cite{maghrebi2017continuous}. To avoid the challenges of observing  spontaneous symmetry breaking in finite volume, in Fig.~\ref{fig:TC_data} we use time-dependent variational principle (TDVP) for infinite matrix product states~\cite{Haegeman} to directly study the properties of the system in the thermodynamic limit. The power-law interactions in Eq.~(\ref{eq:Hamiltonian}) were approximated as a sum of exponential functions and we used bond dimension $\chi=128$ and timestep $\delta t=0.025$ (see~\cite{SOM} for further details).  

We will restrict to states with an infinitely-repeating 2-site unit cell, 
\begin{eqnarray}
\ket{\psi(0)} = \bigotimes_i   \ket{+}_{2i-1}  \left(\cos \phi \ket{+}_{2i} + i\sin \phi \ket{-}_{2i} \right),  \label{eq:initial} 
\end{eqnarray}
where $\phi {=}0$ corresponds to spins polarized along the $x$-axis. The CTC order parameter is defined as 
$\langle \sigma^+ \rangle \equiv (1/2) \sum_i |\langle \sigma_i^+ \rangle|$, i.e.,  we average the \emph{absolute} expectation value of $\sigma^+ {\equiv} (\sigma_x+i\sigma_y)/2$ over the sites in the unit cell. The absolute value of $\sigma^+$ is chosen in order to avoid cancellations due to different relative phases for $\langle\sigma^+_i\rangle$ on different sites of the $2$-site unit-cell. 
We confirm that for $\alpha {\lesssim}  2.5$ the order parameter $\langle \sigma^+ \rangle$ acquires a finite expectation value in the ground state of $D$. The local $h_z$ field drives rotations in the $xy$-plane, causing the order parameter to oscillate periodically -- the anticipated hallmark of the CTC phase.  Fig.~\ref{fig:TC_data} illustrates this by the dynamics of $D(t) \equiv \langle \psi(t) | D|\psi(t) \rangle$ (normalized by the value at $t=0$), the von Neumann bipartite entanglement entropy $S_E(t)$ and the order parameter $\langle \sigma^+(t) \rangle$. 

Fig.~\ref{fig:TC_data} (a)-(c) are for the $x$-polarized ($\phi{=}0$) initial state. As $h_z$ is increased, the CTC  phase is stabilized:  $D$ is well conserved, while $\langle \sigma^+(t) \rangle$ remains constant. For intermediate $h_z$, $D$ does not decay to zero as typically seen in periodically driven systems~\cite{Machado2020}. This is due to the fact that $h_z$ is a parameter in our Hamiltonian, rather than a driving frequency which pushes the system to infinite temperature.  The fact that $\langle \sigma^+(t) \rangle$ remains approximately constant implies periodic oscillations in $\sigma_x(t)$ and $\sigma_y(t)$ with a period $T \approx 2\pi/h_z$. Due to the asymmetry between $J_x$ and $J_y$ couplings, $\langle \sigma^+(t) \rangle$ is not exactly conserved over time even in the prethermal phase, instead it oscillates between maxima (minima) when pointing along the $x$- or $y$-axis. This is also the cause of the small oscillations observed in $D$ on the prethermal plateau. As our chosen initial state $\ket{\psi(0)}$ is close to the ground state of $H_\mathrm{eff}$ (but mid-spectrum for $H$), the growth of $S_E(t)$ is strongly suppressed for large $h_z$. 

At high temperatures, the effective Hamiltonian transitions out of the CTC  phase to a trivial disordered phase. The impact of energy density on the dynamics can be seen by varying $\phi$ in Eq.~(\ref{eq:initial}) to increase the energy density of the initial state. Dynamics for various choices of $\phi$ can be seen in Fig.~\ref{fig:TC_data} (d)-(f). These states are spread through the spectrum of $H_\mathrm{eff}$, with $D(0)/N \approx\{-0.35,-0.26, -0.15 -0.05, 0.05\}$ respectively. For all these states, $D$ is well conserved, thus we remain in a prethermal phase. However, the increase in energy density means that the prethermal Gibbs state eventually becomes a high-temperature state and CTC order is lost. This is accompanied by $\langle \sigma^+(t)\rangle $ decaying to zero and faster growth of $S_E(t)$. 

\begin{figure}[tbh]
    \centering
    \includegraphics[width=0.49\textwidth]{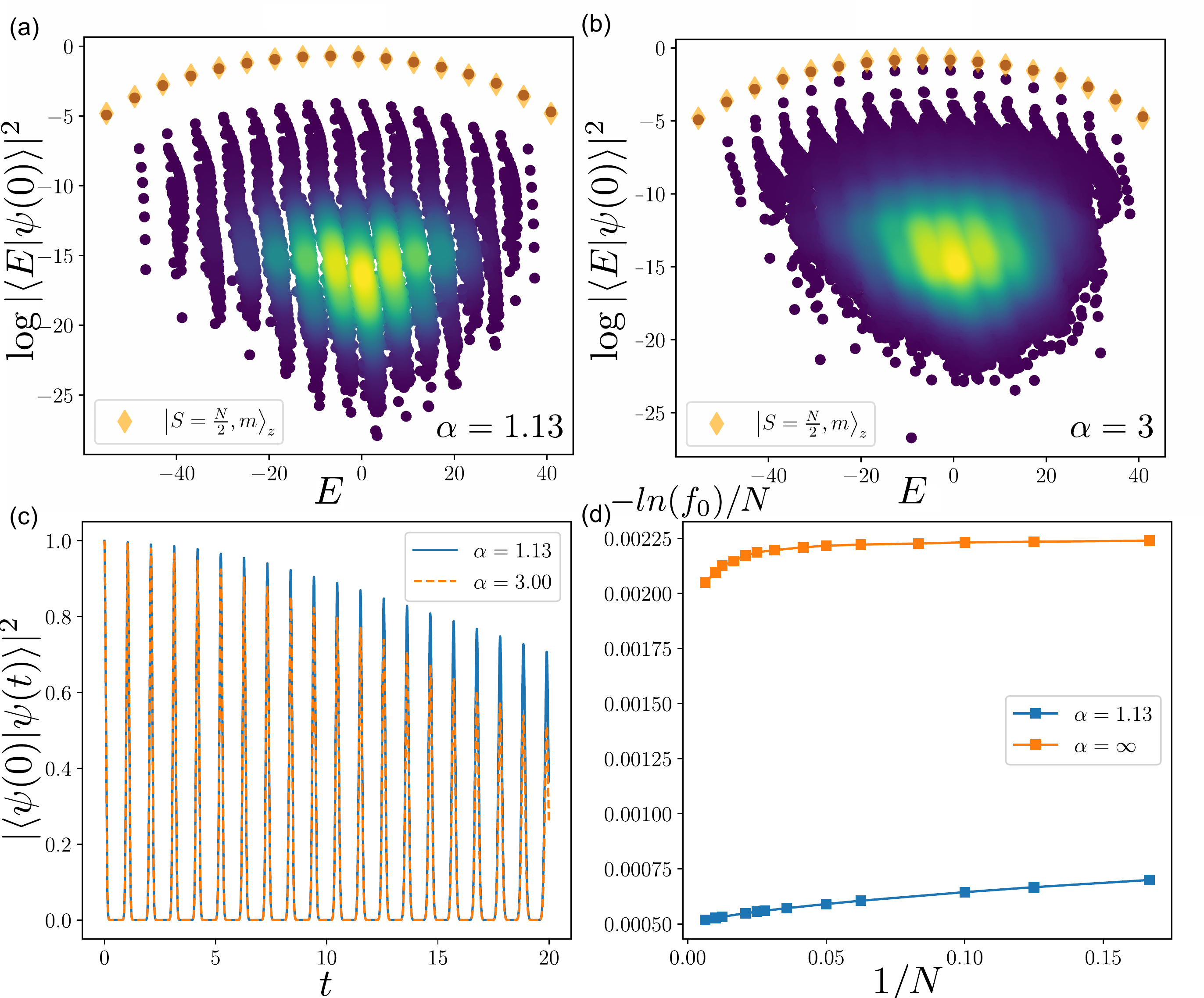}
    \caption{Quantum many-body scars near the isotropic limit of the model in Eq.~(\ref{eq:Hamiltonian}), with $J_x{=}-0.8$, $J_y{=}-1$, $J_z {=}-0.95$, $h_z=3$ and system size $N=16$.
    The initial state $\ket{\psi(0)}$ is $x$-polarized [$\phi{=}0$ in Eq.~(\ref{eq:initial})]. 
    Top row: Eigenstate overlap with $\ket{\psi(0)}$ for both long-range (a) and short-range (b) models. In both cases, the top band of eigenstates are the QMBS  eigenstates, which are well approximated by maximal spin su(2) basis states in the $z$-direction, denoted by diamonds. (c): Quantum fidelity revivals from the initial state $|\ket{\psi(0)}$, for both long and short range model. (d): Finite-size scaling of the fidelity density $-\ln(f_0)/N$, where $f_0$ is the height of the first fidelity revival. The fidelity density was obtained using finite-TDVP with varying system size, bond dimension $\chi=300$ and timestep $\delta t=0.02$. 
    }
    \label{fig:xyz_scars}
\end{figure}
{ \bf \em Many-body scars via ``tunnels-to-towers".---} Close to the isotropic point, $J_x{=}J_y{=}J_z$, we find QMBS arise in the model (\ref{eq:Hamiltonian}) due to an approximate ``tunnels to towers"  mechanism~\cite{Dea2020}. The field term in the Hamiltonian in Eq.~(\ref{eq:Hamiltonian}), $Z=\sum_j \sigma_j^z$, possesses a spectrum generating algebra with respect to the raising operator of the standard su(2) representation, 
$    \left[Z , \sigma^+ \right] = 2 \sigma^+$.
This trivially guarantees the eigenstates of $Z$ are equidistant `towers'. Taking $Z$, one can form a Hamiltonian by adding some additional term, specially chosen so as to preserve only a single tower of eigenstates of $Z$ as eigenstates of the full Hamiltonian, while generically mixing other towers such that the resulting model is non-integrable~\cite{Dea2020}. The preserved tower of eigenstates are found to be QMBS eigenstates. For example, they have subthermal entanglement entropy and coherent dynamics in all observables can be witnessed by preparing initial states with dominant support on the scarred subspace.  Previous constructions of scarred Hamiltonians of this form have preserved a single tower of eigenstates exactly, in the sense that they remain exact eigenstates of the full Hamiltonian and therefore remain equidistant in energy. Sufficiently close to the isotropic point of the Hamiltonian in Eq.~(\ref{eq:Hamiltonian}), these conditions are satisfied \emph{approximately} (in~\cite{SOM} we quantify this). In this sense, a set of QMBS  eigenstates are found in the spectrum of the Hamiltonian, which are approximately equidistant in energy and resemble some subset of exact eigenstates of $Z$. These QMBS eigenstates require weakly broken su(2) symmetry and their presence is largely independent of $\alpha$.

In Fig.~\ref{fig:xyz_scars} we demonstrate the existence of QMBS eigenstates by exact diagonalization of a $N{=}16$ site chain. We consider couplings close to the isotropic point, $J_x{=}{-}0.8$, $J_y{=}{-}1$, $J_z{=}{-}0.95$. In Figs.~\ref{fig:xyz_scars}(a)-(b) We plot the overlap of eigenstates with the $x$-polarized state [$\phi{=}0$ in Eq.~(\ref{eq:initial})], for both long-range ($\alpha{=}1.13$) and short-range ($\alpha{=}3$) models. In both cases, we see a top band of scarred eigenstates and note they resemble the large spin su(2) basis states in the $z$-direction, $\vert S{=}N/2,m\rangle$. These are precisely the eigenstates of $Z$ which are approximately preserved as eigenstates of the full Hamiltonian. We note that in sectors with smaller total-$S$ the towers of $Z$ eigenstates no longer accurately describe the eigenstates of the full Hamiltonian (e.g., for the N\'eel state in the $x$-direction there are no visible towers).
As the $x$-polarized state has dominant support on the QMBS eigenstates which are approximately equidistant in energy, it follows that initializing the system in this state results in a periodic trajectory in the Hilbert space and revivals in the many-body wavefunction, demonstrated by the revivals in quantum fidelity, $f(t)=\vert \psi(0) \vert \psi(t) \rangle \vert^2$, in Fig.~\ref{fig:xyz_scars}(c). We confirm that the non-ergodicity in the dynamics from such initial states persists in the thermodynamic limit by performing finite-size scaling of the fidelity density $-\ln(f_0)/N$, where $f_0$ is the amplitude of the first fidelity revival. The fidelity density in Fig.~\ref{fig:xyz_scars}(d) is found to converge to a value much smaller than $\ln 2$, expected for a random initial state in a thermalizing system. The extrapolated value is of the same order for both long- and short-range models, indicating the persistence of ergodicity breaking due to QMBS.

\begin{figure}[tbh]
	\includegraphics[width=0.7\linewidth]{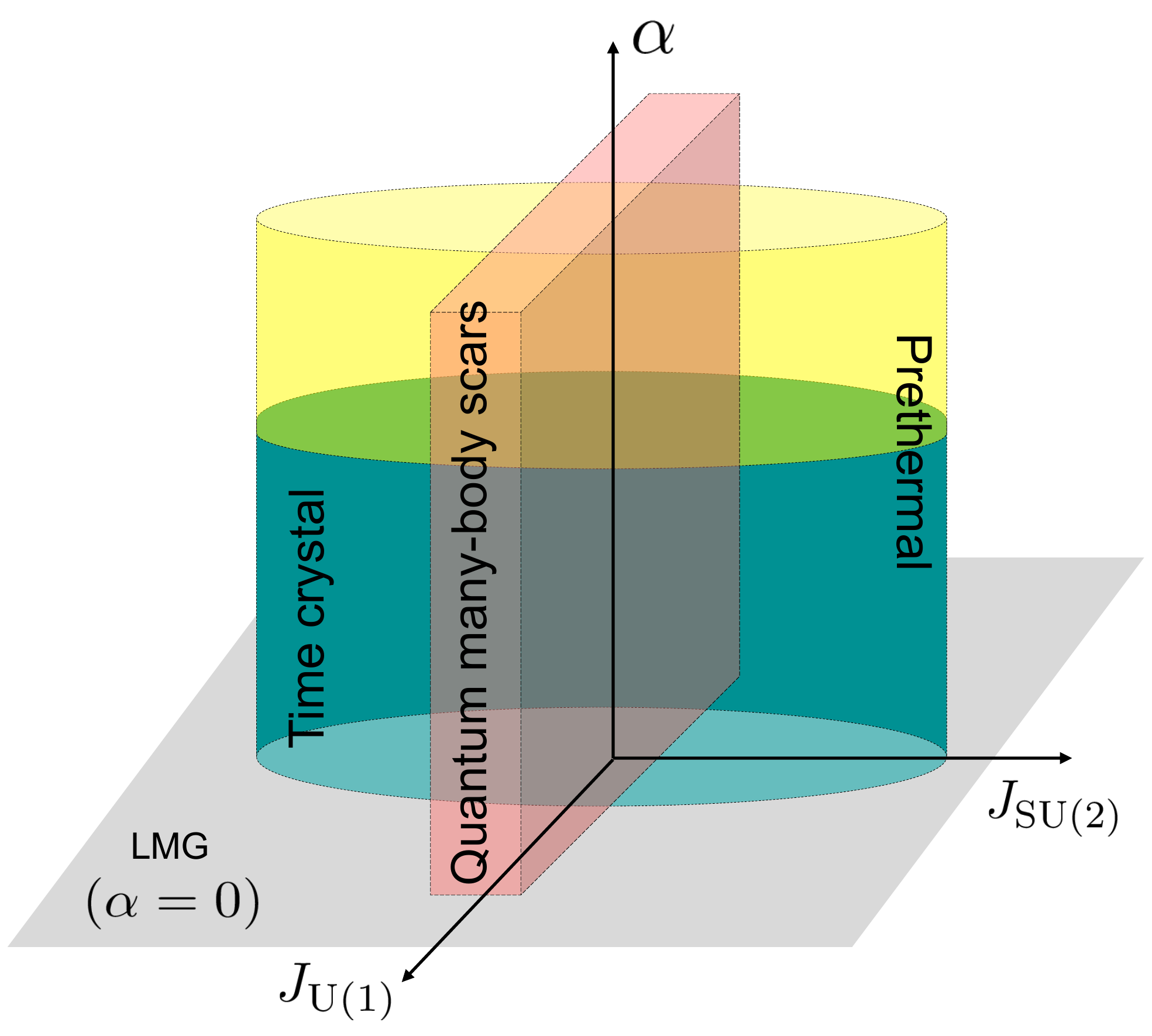}
	\caption{Schematic illustration of the dynamical phase diagram of the model in Eq.~(\ref{eq:Hamiltonian}) as a function of U(1) symmetry breaking $J_\mathrm{U(1)}$, SU(2) symmetry breaking $J_\mathrm{SU(2)}$ and interaction range $\alpha$. Within the prethermal regime (yellow), CTC phase emerges for small $\alpha$ and small $J_\mathrm{U(1)}$ (green). QMBS (red) are independent of $\alpha$ but require small $J_\mathrm{SU(2)}$. 
	We also indicated the solvable LMG limit ($\alpha{=}0$) shown in grey. 
	}\label{fig:summary}
\end{figure}

{ \bf \em Dynamical phase diagram.---}
 Fig.~\ref{fig:summary} is a schematic summary of our numerical investigation of the phase diagram as a function of $\alpha$, anisotropy that leads to U(1) symmetry breaking, $J_\mathrm{U(1)} {\equiv} |J_x-J_y|$, and SU(2) symmetry breaking in the rotating frame, $J_\mathrm{SU(2)} {\equiv} |(J_x+J_y)/2-J_z|$ (see~\cite{SOM} for numerical data). The field $h_z$ is assumed to be fixed to some large value, $h_z \gg J_\nu$, and the remaining dependence on $J_\nu$ and $\alpha$ is sketched. We the discuss three main regions of this phase diagram below: (i) the prethermal regime, (ii) CTC phase, and (iii) QMBS. 

(i) The static prethermal theorem~\cite{Abanin2017} shows that a Hamiltonian of the form $H=H_0 {+} h N$, with $N$ possessing an integer spectrum, in the limit of large $h$ can be brought into a form $ D {+} V {+} h N$ through a series of unitary rotations, where $D$ commutes with $N$ and $V$ is an exponentially small correction in $h$. Thus, for exponentially long times in $h$, the dynamics is governed by an effective prethermal Hamiltonian $H_\mathrm{eff}=D {+} h N$, which has a U(1) symmetry generated by $N$.  In~\cite{SOM} we explicitly perform the unitary rotation to first order, finding the correction terms in $V$ contain an inverse power law dependence on $\alpha$, similar to the original Hamiltonian, with the overall prefactors $J_\mathrm{U(1)}^2/h_z$, $J_\mathrm{U(1)} J_\mathrm{SU(2)}/h_z$. It follows that the prethermal phase is robust, provided $J_\mathrm{U(1)}, J_\mathrm{SU(2)} \ll h_z$.  Moreover, for fixed $h_z$, the prethermal region of the phase diagram takes the shape of an elliptic cylinder as the first-order correction terms have stronger dependence on $J_\mathrm{U(1)}$ than on $J_\mathrm{SU(2)}$, see Fig.~\ref{fig:summary}.

(ii) The CTC phase must be a subset of the prethermal region where the emergent U(1) symmetry of the effective Hamiltonian is spontaneously broken. Due to the  Mermin-Wagner theorem, in 1D this can only happen if the interactions are sufficiently long-ranged~\cite{Machado2020,maghrebi2017continuous}. Consistent with this, we observe a transition when $\alpha {\sim} 2.5 $ from a trivial U(1)-preserving phase to a CTC phase. Thus, we expect the prethermal CTC phase to exists within the bounded cylindrical region depicted in Fig.~\ref{fig:summary}. 

(iii)  
The robustness of QMBS is  determined by how well the interacting terms approximately preserve a single tower of eigenstates of $Z$, which is solely dependent on the model's proximity to the fully isotropic point, $J_x{=}J_y{=}J_z$. At the isotropic point the model possesses SU(2) symmetry irrespective of $\alpha$, hence the QMBS region has no $\alpha$-dependence and it is bounded by two planes perpendicular to the $J_\mathrm{SU(2)}$ axis. The boundary is sharp as the QMBS behavior diminishes exponentially with $J_\mathrm{SU(2)}$~\cite{SOM}.

{ \bf \em Conclusions.---} We introduced a long-range interacting XYZ spin model 
that realizes two types of weak ergodicity breaking phenomena -- a CTC phase as well as QMBS states, allowing to controllably  tune between them by varying the interaction couplings. Perhaps the most interesting implication of our study, as indicated in Fig.~\ref{fig:summary}, is that when both $J_\mathrm{U(1)}$ and  $J_\mathrm{SU(2)}$ are small, the model simultaneously hosts QMBS and CTC, raising many interesting questions about their distinction. Our results suggest that the two phenomena can be distinguished by probing the quench dynamics from different initial states. QMBS occur for initial states that have a large overlap with the large-$S$ spin sector (such as the $x$-polarized state), regardless of energy density. The lifetime of the scarring revivals is exponentially sensitive to $J_\mathrm{SU(2)}$. Moreover, QMBS place stronger constraints on the dynamics, leading to the wave function fidelity revivals, in addition to the oscillations of a local order parameter.  In contrast, the CTC manifests for initial states that have low energy density with respect to $D$, but not necessarily large support on a large-$S$ spin sector. Hence, CTC will persist for other initial states, such as the 2-site unit cell states in Eq.~(\ref{eq:initial}), as long as those are below critical energy density with respect to $D$. The CTC oscillations depend weakly on $J_\mathrm{SU(2)}$ but their lifetime is exponentially long in $J_\mathrm{U(1)}/h_z$. In future work, it would be interesting to analyze the behavior of CTC for initial states beyond period-2, e.g., the spiral states recently used in Ref.~\cite{Jepsen2021}, as well as possible realizations of CTC and QMBS in local models in higher dimensions.

{ \bf \em Acknowledgements.---} We thank Alessio Lerose for useful discussions. K.B., A.H., and Z.P. acknowledge support by EPSRC Grants No. EP/R020612/ 1 and and No. EP/M50807X/1, and by the Leverhulme Trust Research Leadership Award No. RL-2019-015. I.M. was supported by the US Department of Energy, Office of Science, Basic Energy Sciences, Materials
Sciences and Engineering Division. Statement of compliance with EPSRC policy framework on research data: This publication is theoretical work that does not require supporting research data.

\bibliography{references}

\onecolumngrid 
\newpage

\section{Supplemental Online Material for ``Tuning between continuous time crystals and many-body scars in long-range XYZ spin chains"}
\begin{center}
    Kieran Bull$^1$, Andrew Hallam$^1$, Zlatko Papi\'c$^1$, and Ivar Martin$^2$ 
    
    \vspace*{0.2cm}

    {\footnotesize \sl $^1$School of Physics and Astronomy, University of Leeds, Leeds LS2 9JT, United Kingdom, \\
    $^2$Material Science Division, Argonne National Laboratory, Argonne, IL 08540, USA}
\end{center}

{
 \small In this Supplementary Material, we derive the first order corrections to the prethermal Hamiltonian and give some background on the towers-to-tunnel construction of quantum many-body scars. We provide further details on the numerical simulations used in the main text, including the extensive numerical investigation of the phase diagram.   
}
\vspace*{0.4cm}

\setcounter{subsection}{0}
\setcounter{equation}{0}
\setcounter{figure}{0}
\renewcommand{\theequation}{S\arabic{equation}}
\renewcommand{\thefigure}{S\arabic{figure}}
\renewcommand{\thesubsection}{S\arabic{subsection}}

\twocolumngrid

\section{First-order correction to the prethermal Hamiltonian}

The static prethermal theorem~\cite{Abanin2017} shows that a Hamiltonian of the form $H=H_0 + \nu N$, where  $N$ has integer spectrum, can be transformed into a form $ D+V+ \nu N$, where  $D$ commutes with $N$ and $V$ is an exponentially small correction in $\nu$. This implies that for exponentially long times dynamics is governed by an effective prethermal Hamiltonian $H_\mathrm{eff}=D+\nu N$, which has a U(1) symmetry generated by $N$. Here we apply this construction to the long-range Hamiltonian introduced in the main text and examine the leading order correction to the prethermal behavior.

The construction of $H_\mathrm{eff}$ proceeds recursively via a sequence of unitary rotations. Here we explicitly evaluate the rotation to first order. We make use of the same notation introduced in the main text,  $J_\mathrm{U(1)} \equiv J_x - J_y$ and $J_\mathrm{SU(2)} \equiv (J_x+J_y)/2 -J_z$. Our starting point is the Hamiltonian $H$:
\begin{eqnarray}
\notag H &=&  \underbrace{\sum_{i>j} \frac{J_x}{\vert i-j\vert^{\alpha}} \sigma_i^x \sigma_j^x + \frac{J_y}{\vert i-j\vert^{\alpha}} \sigma_i^y \sigma_j^y + \frac{J_z}{\vert i-j\vert^{\alpha}} \sigma_i^z \sigma_j^z}_{H_0} \\
&+& \underbrace{h_z \sum_{i} \sigma_i^z}_{\nu N},
\end{eqnarray}
where we have denoted the field term by $\nu N$. Following Ref.~\cite{Abanin2017}, we define 
\begin{eqnarray}
    D_n &=& \frac{1}{T} \int_0^T e^{ i t \nu N} H_n e^{-i t \nu N} dt,\\
    V_n &=& H_n - D_n, \\
    A_n &=& -\frac{i}{T} \int_0^T \int_0^t e^{i s \nu N} V_n e^{-i s \nu N} ds dt,
\end{eqnarray}
where $T= 2 \pi/\nu$. These allow us obtain the $n$th order Hamiltonian from 
\begin{eqnarray}\label{eq:rec}
        \nu N + H_{n+1} &=& e^{-A_n} (\nu N + H_n) e^{A_n}.
\end{eqnarray}
For our particular model, we have
\begin{eqnarray}
    D_0 &=& \sum_{i>j} (J_\mathrm{SU(2)}+J_z) \frac{\sigma_i^x \sigma_j^x + \sigma_i^y \sigma_j^y}{ \vert i-j \vert^{\alpha}} + J_z \frac{\sigma_i^z \sigma_j^z}{\vert i-j\vert^\alpha}, \;\;\;\;\\
    V_0 &=& \frac{J_\mathrm{U(1)}}{2} \sum_{i>j} \frac{\sigma_i^x \sigma_j^x - \sigma_i^y \sigma_j^y}{\vert i - j \vert^\alpha}, \\
    A_0 &=& -\frac{i}{2} \frac{J_\mathrm{U(1)}}{4 h_z} \sum_{i>j} \frac{\sigma_i^x \sigma_j^y + \sigma_i^y \sigma_j^x}{\vert i - j \vert^\alpha}.
\end{eqnarray}
These indeed obey the relation $H_0 = D_0 + V_0$ and, moreover, $[A_0, \nu N] = V_0$. At zeroth order, $V_0$  is the non commuting term responsible for the breaking of the emergent U(1) symmetry at late times (for large $\nu = h_z \gg J$, where $J$ is the largest of $J_x$, $J_y$, $J_z$ couplings) and it has a coefficient proportional to $J_\mathrm{U(1)}$. Therefore, we expect $J_\mathrm{U(1)}$ to have a significant impact on the lifetime of the prethermal phase.

To determine the impact of $J_\mathrm{U(1)}$, $J_\mathrm{SU(2)}$ on the lifetime of the prethermal phase, we must construct higher order correction terms. From the recursion relation Eq.~(\ref{eq:rec}):
\begin{align}
    \nu N + H_1 &= e^{-A_0} (\nu N + H_0) e^{A_0} \nonumber \\
    &\approx(1-A_0)(\nu N + H_0)(1+A_0) \nonumber \\
    &= \nu N + D_0 + V_0 + \underbrace{[\nu N,A_0]}_{-V_0} + [H_0,A_0] \nonumber \\
    &- \nu A_0 N A_0 - A_0 H_0 A_0. 
\end{align}
Thus we find $H_1$ takes the following form:
\begin{align}
    H_1 &= D_0 + \underbrace{[H_0,A_0]}_{\propto 1/h_z} - \underbrace{\nu A_0 N A_0}_{\propto 1 / h_z} - \underbrace{A_0 H_0 A_0}_{\propto 1/h_z^2}.
\end{align}
Explicitly,  
\begin{align}
    H_1 = & D_0 \nonumber \\
    &+ \frac{J_{U(1)}^2}{4 h_z} \sum_{n,m,n \neq m} \frac{\sigma^z_m}{ \vert n - m \vert^{2\alpha}} \nonumber \\
    &+ \frac{J_{U(1)}}{4 h_z} \sum_{\substack{i,n,m \\ i\ \neq n \\ n \neq m}} \frac{(J_x-J_z) \sigma^x_n \sigma^z_m \sigma^x_i - (J_y-J_z)\sigma^y_n \sigma^z_m \sigma^y_i}{\vert i - m \vert^{\alpha} \vert n - m \vert^\alpha} \nonumber \\
    &+ \frac{J_{U(1)}^2}{64 h_z} \sum_{\substack{i,n,m,u,v \\ n \neq m \\ u \neq v}} \frac{(\sigma^x_n \sigma^y_m + \sigma^y_n \sigma^x_m) \sigma^z_i (\sigma^x_u \sigma^y_v + \sigma^y_u \sigma^x_v)}{\vert n - m \vert^\alpha \vert u - v \vert^\alpha} \nonumber \\
    &+ O(1/h_z^2).
    \end{align}
This can be refactored into commuting and non-commuting parts:
\begin{eqnarray}
H_1 = D_1 + V_1    
\end{eqnarray}
where $[D_1, N] = 0, [V_1,N] \neq 0$. For the commuting part $D_1$, one finds:
\begin{align}
    D_1 &= D_0 \nonumber \\
    &+ \frac{J_\mathrm{U(1)}^2}{4 h_z} \left(\sum_{\substack{n,m \\ n \neq m}} \frac{\sigma^z_m}{\vert n - m \vert^{2 \alpha}} + \sum_{\substack{i,n,m \\ i \neq n,m \\ n \neq m}} \frac{\sigma_n^+ \sigma^z_m \sigma_i^- + \sigma_n^- \sigma^z_m \sigma_i^+}{ \vert i - m \vert^\alpha \vert n - m \vert^\alpha} \right) \nonumber \\
    &+ \frac{J_\mathrm{U(1)}^2}{64 h_z} \sum_{\substack{i,n,m,u,v \\ n \neq m \\ u \neq v}} \frac{\sigma_n^+ \sigma_m^+ \sigma^z_i \sigma_u^- \sigma_v^- + \sigma_n^- \sigma_m^- \sigma^z_i \sigma_u^+ \sigma_v^+}{\vert n - m \vert^\alpha \vert u - v \vert^\alpha}. 
    \end{align}
For the non-commuting error term $V_1$, one finds:
\begin{align}
V_1 &= \frac{J_\mathrm{U(1)}}{2 h_z} J_\mathrm{SU(2)} \sum_{\substack{i,n,m \\ i \neq n,m \\ n \neq m}} \frac{\sigma_n^+ \sigma^z_m \sigma_i^+ + \sigma_n^- \sigma^z_m \sigma_i^-}{\vert i - m \vert^\alpha \vert n - m \vert^\alpha}  \nonumber \\
&- \frac{J_\mathrm{U(1)}^2}{64 h_z} \sum_{\substack{i,n,m,u,v \\ n \neq m \\ u \neq v}} \frac{\sigma_n^+ \sigma_m ^+ \sigma^z_i \sigma_u^+ \sigma_v^+ + \sigma_n^- \sigma_m^- \sigma^z_i \sigma_u^- \sigma_v^-}{\vert n - m \vert^\alpha \vert u - v \vert^\alpha}.
 \end{align}
 Crucially, we see the coefficients present in $V_1$ -- the non-commuting error term responsible for destroying the emergent U(1) symmetry with respect to $N$ at late times -- are proportional to $J_\mathrm{U(1)}J_\mathrm{SU(2)}$, $J_\mathrm{U(1)}^2$, implying an increase in these quantities results in a shorter-lived prethermal phase.

\section{Quantum many-body scars from ``tunnels-to-towers"}

Many-body quantum scarring is a mechanism for weak violation of the ETH, which suppresses thermalization from \emph{certain} initial states in non-integrable systems. While generic non-integrable many-body systems thermalize to a Gibbs state whose temperature is determined by the energy density of the initial state, scarred systems may instead undergo periodic dynamics when quenched from special initial states. The periodicity of the dynamics is due to the initial state having large support on special, ``scarred" eigenstates, which have approximately the same energy spacing. 

In \emph{exact} scarred models~\cite{Iadecola2019_2,Iadecola2019_3,OnsagerScars},   the  energy spacing between scarred eigenstates is precisely equal. Therefore, initial states which are an arbitrary superposition of the scarred eigenstates will display perfect oscillations in local observables due to a perfectly reviving wave-function fidelity, $f(t) = \vert \langle \psi(t) \vert \psi(0) \rangle \vert^2$. However, there may not exist a particularly simple initial state that is a superposition of scarred eigenstates, e.g., as in in the Affleck-Kennedy-Lieb-Tasaki (AKLT) model~\cite{BernevigEnt, MotrunichTowers, MoudgalyaHubbard}. In contrast, \emph{approximate} scarred models~\cite{Turner2017,Bull2019,bosonScars,Moudgalya2019,Surace2019} host non-thermal eigenstates which are only approximately equidistant in energy. This can still result in nearly periodic dynamics and suppressed thermalization from special initial states, with  local observables exhibiting decaying oscillations. This scenario is believed to have been observed in experiment~\cite{Bernien2017}, where oscillations of the number of domain walls have been detected in a 51-atom Rydberg simulator prepared in a product (N\'eel) state of atoms.

In models that exhibit equidistant energy eigenstates, either exact or approximate, the Hamiltonian $H$ is typically found to possess an underlying structure  analogous to a spectrum generating algebra (SGA):
\begin{equation}\label{eq:SGA}
    [H,Q^+] = \omega Q^+,
\end{equation}
where $Q^+$ is some raising operator. If a Hamiltonian possesses the SGA in Eq.~(\ref{eq:SGA}), it trivially follows that for every eigenstate that is not annihilated by $Q^+$, there exists a tower of equidistant eigenstates with energy separation $\omega$. For example, take an eigenstate $\ket{E}$, such that $H \vert E \rangle = E \vert E \rangle $. Then we have
\begin{eqnarray}
    H Q^+ \vert E \rangle = (Q^+ H + \omega Q^+) \vert E \rangle  =  (E+\omega)Q^+ \vert E \rangle.    \;\;\;\;
\end{eqnarray}

More non-trivially, Eq.~(\ref{eq:SGA}) may be satisfied only when we restrict to a \emph{subspace} of the Hilbert space -- this is known as a ``restricted" SGA (RSGA)~\cite{MotrunichTowers,MoudgalyaHubbard}. For an eigenstate $\vert \psi_0\rangle$, $H \vert \psi_0 \rangle = E_0 \vert \psi_0 \rangle$, typically a ground state of $H$, 
we then have 
\begin{align}
[H,Q^+] \vert \psi_0 \rangle &= \omega Q^+ \vert \psi_0 \rangle, \quad
[[H,Q^+],Q^+] = 0.
\end{align}
The above properties guarantee the presence of a single tower of equidistant eigenstates, $\vert \psi_n \rangle \equiv (Q^+)^n\vert \psi_0\rangle$. If the Hamiltonian $H$ is non-integrable, but engineered such that the states $\vert  \psi_n\rangle$ are non thermal, the latter states are exact scarred eigenstates.

There exists a specific construction dubbed ``tunnels-to-towers"~\cite{Dea2020}, which produces non-integrable Hamiltonians possessing an RSGA for which the tower of equidistant eigenstates are non-thermal.  The construction relies on an operator $V$, which possess an SGA with respect to some raising operator $Q^+$. One then forms a Hamiltonian $H=H_0 + V$, where $H_0$ is specifically chosen to annihilate a single tower of eigenstates of $V$, while acting like a generic, non-integrable Hamiltonian on all other towers, thus preserving a single tower of eigenstates of $V$ as eigenstates of the full Hamiltonian $H$. Formally, this is summarised as
\begin{align*}
    [V,Q^+] = \omega Q^+, \quad 
    V \vert \psi_0 \rangle = E_0 \vert \psi_0 \rangle, \quad
    H_0  \vert \psi_n \rangle &= 0, 
\end{align*}
where the last condition is valid for all $n$. In Ref.~\onlinecite{Dea2020}, $V$ was chosen to be some generator of a non-Abelian (or $q$-deformed) symmetry group, thus the SGA emerges due to the root structure of the symmetry group's associated Lie algebra.

Now, consider lifting the restriction that $H_0$ completely annihilates a single tower of eigenstates of $V$. One can engineer approximate scarred models if the action of $H_0$ on the states $\vert \psi_n \rangle$, which satisfy $V  \vert \psi_n\rangle  = (E_0 + n \omega) \vert \psi_n\rangle$, is sufficiently close to a projector:
\begin{align}\label{eq:approxT2T}
    H_0 \vert \psi_n\rangle  = \epsilon_0 \vert \psi_n\rangle + \delta_n \vert \psi^{\perp}_n \rangle, \quad \delta_n \ll 1.
\end{align}
From this, we have 
\begin{eqnarray}
&& \langle \psi_n \vert H \vert \psi_n \rangle 
                            = \epsilon_0 + E_0 + n \omega,  \\
\notag && \langle \psi_n \vert H^2 \vert \psi_n \rangle -  \langle \psi_n \vert H \vert \psi_n \rangle ^2 \label{eq:scarVar} = \delta_n \langle \psi_n\vert  (H_0 + V) \vert \psi_{n}^{\perp} \rangle = \vert \delta_n \vert^2. \\
\end{eqnarray}
Here $\vert \psi_n\rangle$ are exact, equidistant eigenstates of $V$, and $H_0$ only weakly mixes these states with the rest of the Hilbert space. Thus, the full Hamiltonian $H{=}H_0+V$ will contain scarred eigenstates which are approximately equidistant in energy and are well approximated by the original tower $\vert \psi_n\rangle$.

\section{Quantum many-body scars in the long-range XYZ model}

The model given in Eq.~(1) of the main text is non integrable. However, close to the isotropic point ($J_x=J_y=J_z$), this model satisfies the conditions of the approximate version of ``tunnels to towers" construction in Eq.~(\ref{eq:approxT2T})), with $V$ being the Zeeman term responsible for the spectrum generating algebra, while $H_0$ acts trivially on the following set of states $\vert n\rangle$:
\begin{eqnarray}
\notag    \vert n \rangle &\equiv& \vert S=\frac{N}{2},-S + n \rangle \\
    &=& \frac{1}{\sqrt{ \begin{pmatrix} N \\ n \end{pmatrix} }} (Q^+)^n \vert 000...\rangle, \quad    Q^+ = \sum_n \sigma_n^+.
\end{eqnarray}
Thus, near the isotropic point, the Hamiltonian contains a set of $N$ scarred eigenstates that are approximately described by the states $\vert n \rangle$. The extent to which the eigenstates of $H$ are characterized by $\vert n \rangle$ is dependent on $\delta_n$, Eq.~(\ref{eq:approxT2T}), which is exactly zero when $J_x=J_y=J_z$. Thus the anisotropy directly controls the robustness of the scars.

 \begin{figure}[ht]
     \centering
     \hspace*{-0.7cm}
     \includegraphics[width=1.2\linewidth]{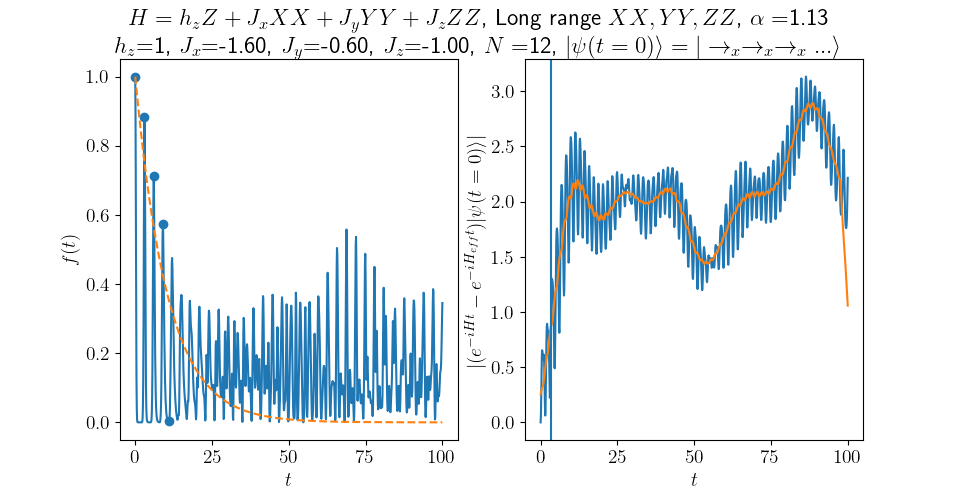}
     \caption{Order parameters for phase identification are computed from the dynamics data shown here. Left: $\tau_\mathrm{f}$ quantifies the robustness of scarring (longer time implies more long-lived oscillations) and it is determined by fitting $e^{-t/\tau_\mathrm{f}}$ to the first five peaks in quantum fidelity time series, $f(t)$. Right: $\tau_\mathrm{U(1)}$ is determined from the distance between trajectories generated by the original Hamiltonian $H$ and the effective prethermal Hamiltonian with U(1) symmetry, $H_\mathrm{eff}$. $\tau_{U(1)}$ is the time for which the distance between these two evolutions, $d(t)$, exceeds $0.5$, see Eq.~(\ref{eq:dist}). }
     \label{fig:tau_dynamics}
 \end{figure} 
\section{Kac normalization of the Hamiltonian}

In the Hamiltonian defined in Eq.~(1) of the main text, the long-range interactions were rescaled by a factor $\frac{1}{\mathcal{N}}$ where $\mathcal{N}$ is the so-called \emph{Kac norm}, defined as
\begin{equation}
    \mathcal{N}=\frac{1}{N-1} \sum_{i \neq j} \frac{1}{\vert i-j \vert^{\alpha}}.
\end{equation}
The Kac norm ensures that the energy-density of the system remains intensive for arbitrary $\alpha$ and makes the it possible to consistently compare the dynamics of the system at different system sizes. All the results in the main text include this normalization factor.

 \begin{figure*}[!h]
     \centering
     \includegraphics[width=0.85\textwidth]{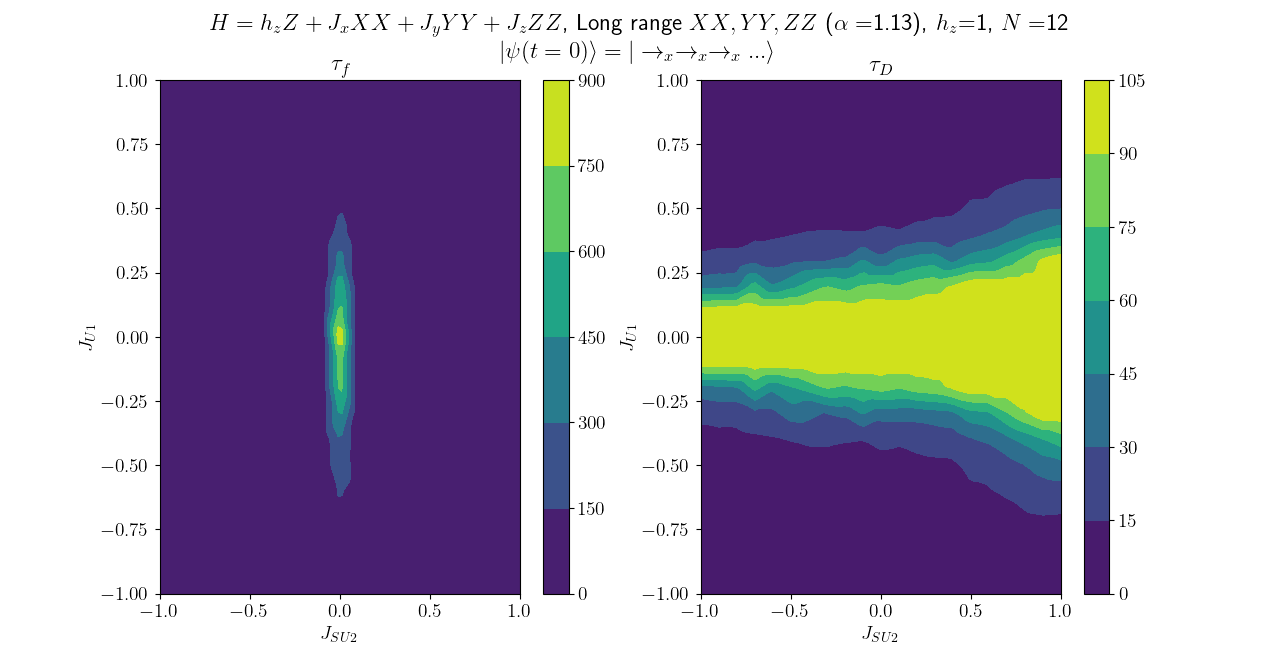}
     \includegraphics[width=0.85\textwidth]{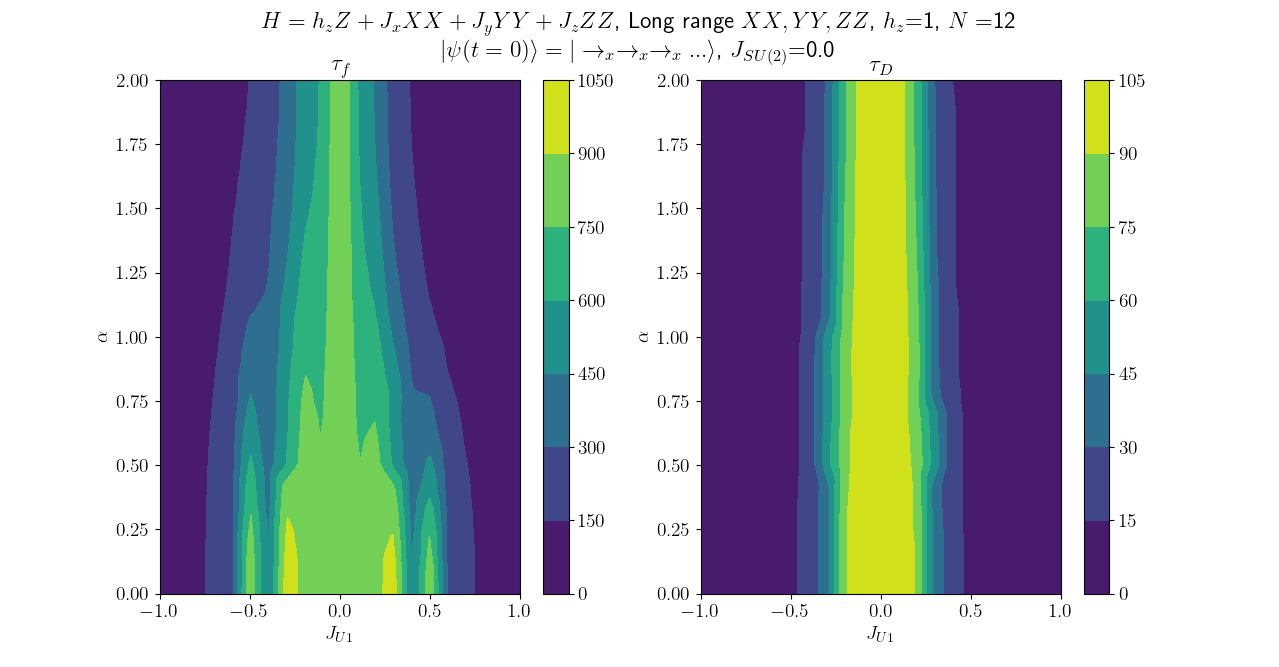}
     \includegraphics[width=0.85\textwidth]{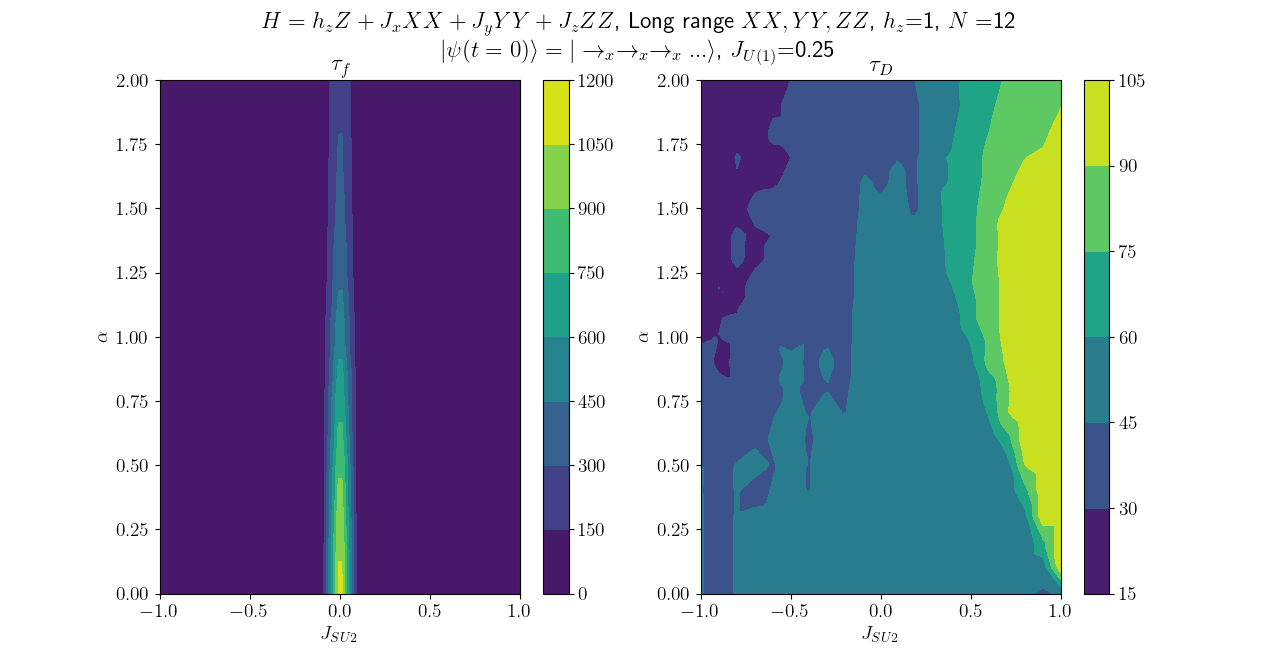}
     \caption{Numerical results for phase diagram slices. Color scale in the left column indicates $\tau_{f}$ (quality of scarring), whereas the color scale in the right column indicates $\tau_{U(1)}$ (quality of prethermal). With Kac norm included, we see essentially no $\alpha$ dependence on $\tau_\mathrm{U(1)}$, thus verifying prethermal theorem  holds for a long-range model with 2-local terms. }
     \label{fig:phaseSlices}
 \end{figure*}
 \section{Numerical study of the phase diagram}\label{sec:numericalPhase}
 
 In the main text we presented a sketch of the dynamical phase diagram of the model in Eq.~(1) as a function of interaction range and anisotropy in the couplings. Here we present results of the numerical analysis behind the sketch, in particular the estimates of the prethermal phase and the quantum many-body scarring regime. To probe these two parts of the phase diagram, we require suitable order parameters for characterizing the relevant physics. To this end, we introduce two timescales: the scarred timescale $\tau_\mathrm{f}$ and the prethermal U(1) timescale $\tau_\mathrm{U(1)}$. We do not explicitly estimate the continuous time crystal regime, as for the present purpose it suffices to know that this phase will be a subset of the prethermal regime, corresponding to sufficiently small $\alpha$ such that spontaneous symmetry breaking is possible (we have estimated this critical $\alpha_c\approx 2.5$ using independent matrix product state calculations in the infinite-size limit, finding agreement with Ref.~\cite{maghrebi2017continuous}). 
 Note that $\tau_\mathrm{f} \rightarrow \infty$ as $J_\mathrm{SU(2)} \rightarrow 0$ and $\tau_\mathrm{U(1)} \rightarrow \infty$ as $J_\mathrm{U(1)} \rightarrow 0$. Thus, we expect the two timescales to provide an appropriate boundary in phase space for the scarred region and the prethermal region, respectively.
 
To estimate $\tau_\mathrm{f}$, we propose to fit an exponential decay $e^{-t/\tau_\mathrm{f}}$ to the first five peaks of the quantum fidelity $f(t)$, yielding the revival decay time $\tau_\mathrm{f}$.  To probe the lifetime of the prethermal phase, we consider the following distance $d(t)$ between the trajectories induced by the bare Hamiltonian and that of the effective prethermal Hamiltonian:
 \begin{eqnarray}\label{eq:dist}
     d(t) = \vert (e^{-i H t} - e^{-i H_\mathrm{eff} t)} \vert \psi(t=0) \rangle \vert , 
 \end{eqnarray}
 where $\vert \psi(t=0) \rangle$ is the polarized state in the $x$-direction. The effective Hamiltonian $H_\mathrm{eff}$ in the rotating frame was defined in the main text.  We denote $\tau_\mathrm{U(1)}$ as the time when $d(t)$ increases above a certain cutoff, which we take to be $d_\mathrm{cut}=0.5$. In practice, we find this quantity is oscillatory, so $\tau_\mathrm{U(1)}$ is the time for which the moving average of $d(t)$ exceeds $d_\mathrm{cut}$. Fig.~\ref{fig:tau_dynamics} shows an example time series for one point in the phase space, and how the above criteria are applied to extract $\tau_f$ and $\tau_\mathrm{U(1)}$.

 Fig \ref{fig:phaseSlices} shows the numerical results for the extracted quantities $\tau_\mathrm{f}$, $\tau_\mathrm{U(1)}$, across various slices through the phase diagram. These results confirm the sketch presented in the main text. We notice some subtle features that were neglected in the sketch: the scarred region ($\tau_\mathrm{f}$) has some $J_\mathrm{U(1)}$ dependence,  and the prethermal region has somewhat stronger $J_\mathrm{SU(2)}$ dependence than depicted in the sketch.

   \begin{figure}[ht]
     \centering
     \includegraphics[width=0.4\textwidth]{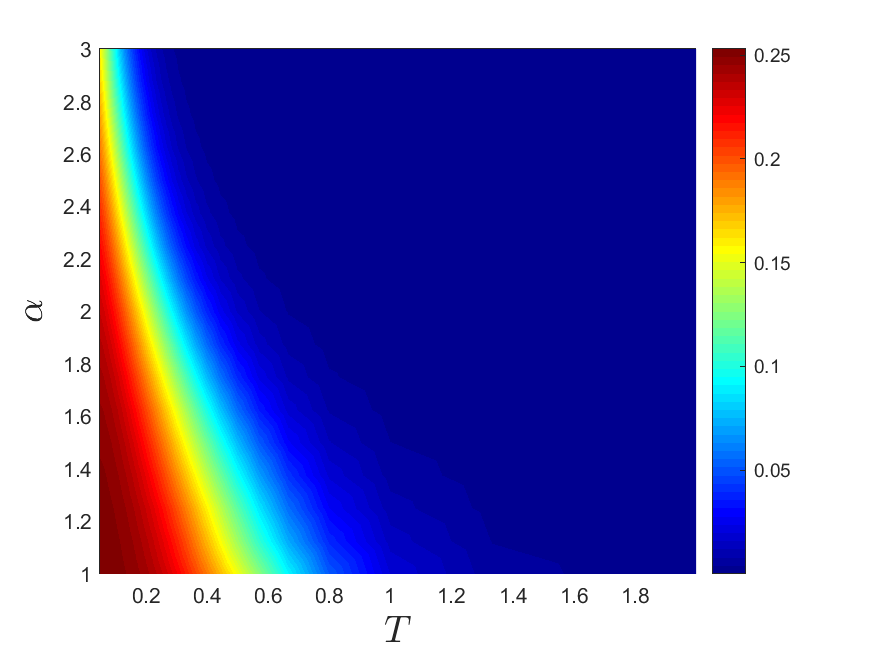}
     \caption{The order parameter $\langle S^+_1S^-_N \rangle$ fxor $J_x=J_y=1.2,J_z=1$, $N=40$ as a function of temperature $T$ and interaction length scale $\alpha$ for the thermal Gibbs state of $D$.}
     \label{fig:Tvsalpha}
 \end{figure}

\section{Temperature and $\alpha$ dependence of time-crystal phase}
 
 It is well known that the ground state of the long-range XXZ model exhibits spontaneous symmetry breaking for sufficiently small $\alpha$~\cite{maghrebi2017continuous}, however we are not aware of any investigation of the stability of this phase at finite temperatures. In order to study the robustness of the U(1) symmetry breaking, we created a thermofield double (TFD) representation of the thermal Gibbs state of $D$,
 \begin{eqnarray}
\rho_\beta \rightarrow \frac{1}{\sqrt{Z}}\sum_n e^{-\beta E_n/2}\ket{n}\otimes\ket{n},
 \end{eqnarray}
where $E_n$ are the energy eigenvalues of $D$ and $\ket{n}$ are its eigenvectors. The TFD state can be calculated using matrix-product state methods by evolving in imaginary time from the infinite temperature state $\ket{\psi(\beta=0)}=\bigotimes \sum_{i=1}^d \ket{i}\otimes\ket{i}$ in steps of $\delta \beta$.
 
 In Fig.~\ref{fig:Tvsalpha} the U(1) symmetry breaking order parameter $\langle S^+_1S^-_N\rangle$ is shown for thermal states at $N=40$ and $J_x=J_y=1.2$, $J_z=1$ as a function of temperature and $\alpha$. For $\alpha<2.5$ there is a well-defined low temperature U(1) symmetry breaking phase which vanishes as $\alpha$ increases to $\alpha_c\approx 3$ and the model transitions to the symmetry-preserving XY-phase, familiar from the nearest neighbour XXZ model. 
 
\section{Time Crystal: Absence of Fidelity Revivals}

\begin{figure}[ht]
     \centering
     \includegraphics[width=0.5\textwidth]{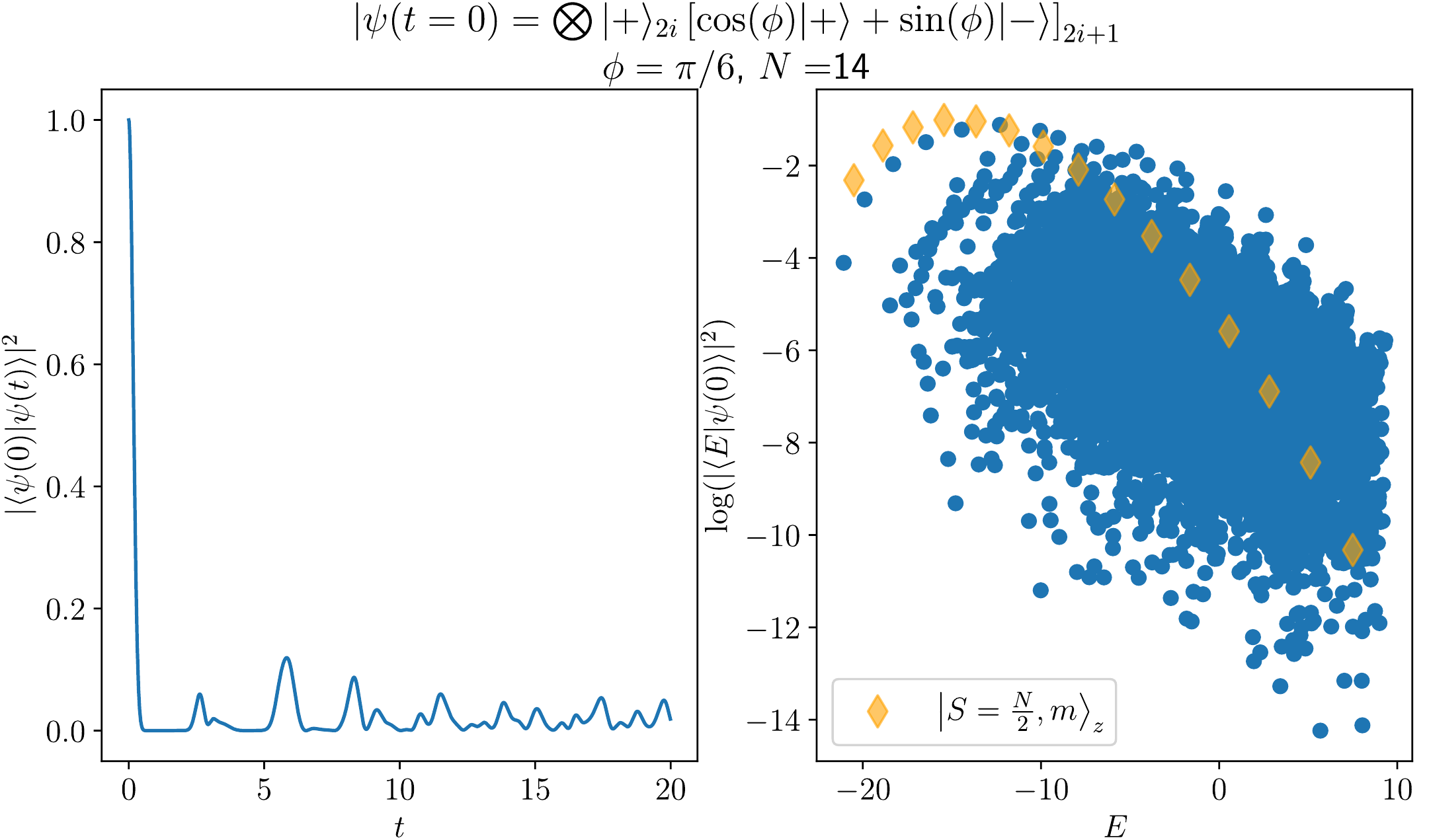}
     \caption{Abscence of fidelity revivals (left) or support on scarred eigenstates (right) for the two-site unit cell initial state at $\phi=\pi/6$. This indicates that the coherent oscillations in a CTC order parameter are not a trivial consequence of fidelity revivals, and the CTC behavior emerges from a mechanism distinct from QMBS. The dynamics were found for the long-range XYZ model at  system size $N=14$ with  $\alpha=1.13$, $J_{x}{=}-0.4$, $J_{y}{=}-2.0$, $J_{z}{=}-1$, $h_z=1.0$.}
     \label{fig:tc_no_fid}
 \end{figure}

Whenever there is a periodic trajectory of the wavefunction in a Hilbert space, quantified by revivals in the the wavefunction fidelity $f(t) = \vert \langle \psi(t) \vert \psi(0) \rangle \vert^2$ (as seen with QMBS), it necessarily follows that any observable will show coherent oscillations. Therefore, in the overlapping region of QMBS and CTC in the phase diagram given in the main text, one may question whether oscillations in the order parameter of the spontaneously broken symmetry (which we claim indicates a CTC) may simply be a consequence of the wavefunction revivals and therefore of 'scarred origin'. We demonstrate the two effects are distinct, even in the overlapping region, by considering the two-site unit cell initial state given in the main text in Eq.~(3) for intermediate $\phi=\pi/6$, for which we have shown the existence of a CTC. This initial state does not possess SU(2) symmetry, and therefore we do not expect it to have support on the scarred eigenstates or exhibit wavefunction revivals, indicating the presence of CTC-like behavior in the spontaneous-symmetry-breaking order parameter should not be of a 'scarred origin'. We demonstrate this is the case for a finite system in Fig \ref{fig:tc_no_fid}. 

\section{Implementation of time-dependent variational principle} 
 
The dynamics results in the main text were obtained using the time-dependent variational principle (TDVP) over matrix-product states (MPS) using the mixed canonical gauge, as described in Ref.~\cite{vanderstraeten2019tangent}. MPS methods cannot exactly describe power-law decaying interactions, instead the power-law can be approximated as a sum of exponential terms, 
\begin{eqnarray}
\frac{1}{|i-j|^\alpha}=\sum^{N_e}_n f_n e^{-\lambda_n (|i-j|}.
\end{eqnarray}
We chose to approximate the power-law using $N_e=8$ exponential function for all data presented in this work.  While the Hamiltonian can be represented as a matrix-product operator, the time-evolution due to each of these exponential interactions can be more efficiently calculated using the method presented in Ref.~\cite{zauner2018variational}. 

\begin{figure}[ht]
     \centering
     \includegraphics[width=0.5\textwidth]{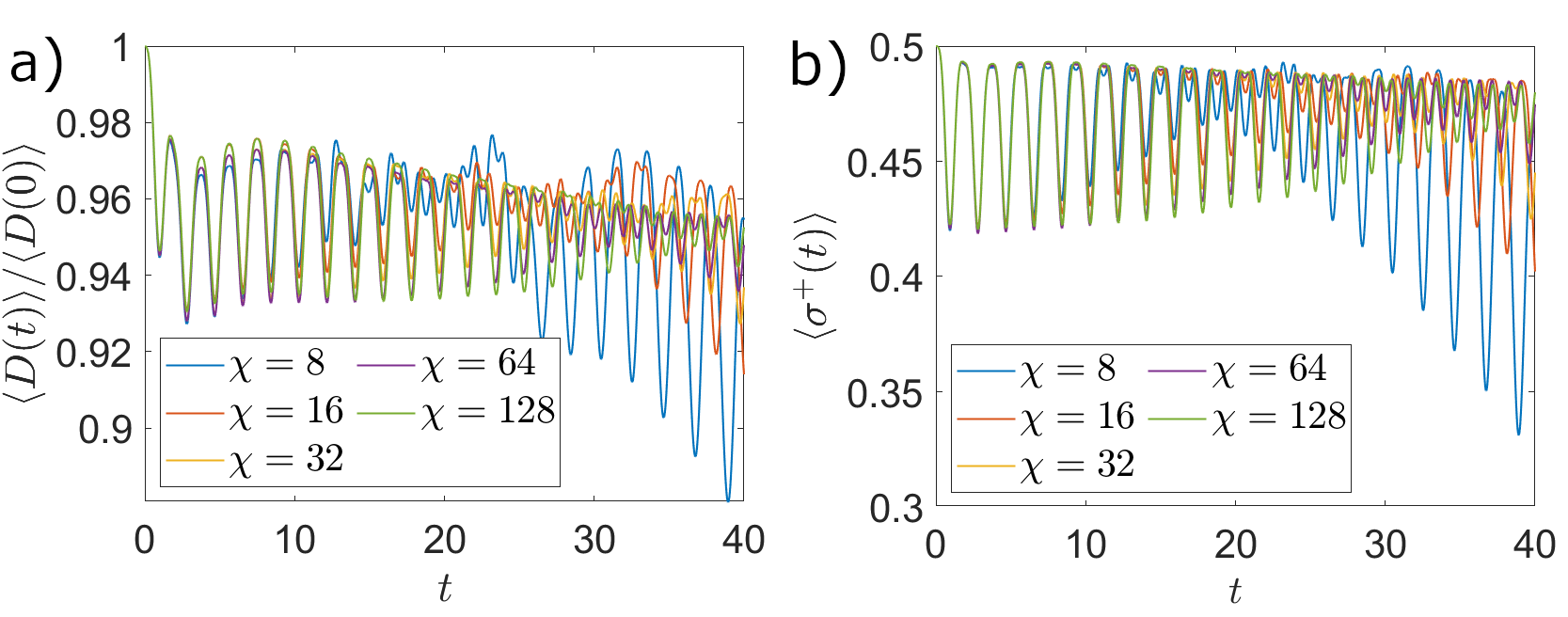}
     \caption{(a) Convergence with bond dimension $\chi$ of the expectation value of the prethermal Hamiltonian $D$ in the time-evolved state. The quantity is normalized by its value at time $t{=}0$. (b) Order parameter $\langle \sigma^+(t) \rangle$ as defined in the main text. 
    All plots are for the infinite long-range XYZ model with  $\alpha=1.13$, $J_{x}{=}-0.4$, $J_{y}{=}-2.0$, $J_{z}{=}-1$, $h_z=0.75$. The initial state is given by Eq.~(\ref{eq:initial}) with $\phi{=}0$. }
     \label{fig:chicomparison}
 \end{figure}

MPS methods capture the exact dynamics of the system provided that the entanglement entropy is not too large. Unfortunately, the entanglement entropy typically grows linearly after a quench, as we see in Fig.~1 of the main text. In Fig.~\ref{fig:chicomparison} we present dynamics of $\langle D(t) \rangle/\langle D(0)\rangle$ and $\langle \sigma^+ (t)\rangle$ at several bond dimensions up to $\chi=128$. At small bond-dimension both of these quantities appear to decay. However,  by increasing $\chi$, we see this decay is an artifact of small bond dimension and the prethermal time crystal becomes robust at large $\chi$. 

\end{document}